\documentclass[jcp,reprint,nofootinbib,twocolumn,amsmath,amsfonts,amssymb]{revtex4}
\usepackage{color}
\usepackage{bm}
\usepackage{graphicx}
\usepackage{array,longtable}
\usepackage{epic,eepic}

\newcommand{\bra}{\langle}
\newcommand{\ket}{\rangle}

\begin{document}
\title{Extended M{\o}ller-Plesset
 perturbation theory for dynamical and static correlations}
\author{Takashi Tsuchimochi}
\email{tsuchimochi@mit.edu} 
\author{Troy Van Voorhis}
\affiliation{Department of Chemistry, Massachusetts Institute of Technology, 77 Massachusetts Ave., Cambridge MA 02139}

\begin{abstract}
We present a novel method that appropriately handles both dynamical and static electron correlation in a balanced manner, using a perturbation theory on a spin-extended Hartree-Fock (EHF) wave function reference. While EHF is a suitable candidate for degenerate systems where static correlation is ubiquitous, it is known that most of dynamical correlation is neglected in EHF. In this work, we derive a perturbative correction to a fully spin-projected self-consistent wave function based on second-order M{\o}ller-Plesset perturbation theory (MP2). The proposed method efficiently captures the ability of EHF to describe static correlation in degeneracy, combined with MP2's ability to treat dynamical correlation effects. We demonstrate drastic improvements on molecular ground state and excited state potential energy curves and singlet-triplet splitting energies over both EHF and MP2 with similar computational effort to the latter.
\end{abstract}
\maketitle
{\it Introduction.}
An efficient and accurate treatment of both dynamical and static electron correlation effects has been elusive in electronic structure theory. Single reference methods such as second-order M{\o}ller-Plesset perturbation theory (MP2) and coupled-cluster singles and doubles (CCSD) achieve high accuracy in computed observables for non-degenerate systems,\cite{Bartlett07} but it is well known that they cannot describe static correlation in degenerate systems. This failure is undoubtedly attributed to the reference wave function: Hartree-Fock (HF). A HF reference is qualitatively inadequate for (nearly-) degenerate systems where the true wave function is multi-determinantal in nature. Complete active space self-consistent field (CASSCF) resolves this problem by treating all the configurations in an active space, yielding a multi-reference state, and usually represents a good starting point when an appropriate active space is chosen. When the residual dynamical correlation is included through a perturbative correction\cite{Hirao92,Andersson94} or configuration interaction (CI), CASSCF can achieve very accurate results both for the ground state and excited states. However, none of these are black-box, and their computational cost is very expensive. 

Yet another approach to tackling static correlation may be spin-extended HF (EHF),\cite{Lowdin55,Mayer80} which is also called spin-projected HF. The idea behind it is to optimize orbitals of a broken symmetry Slater determinant $|\Phi_0\ket$, called a {\it deformed} state, projected by a spin-projection operator $\hat P$ so that the total energy of the projected state,
\begin{align}
E_{\rm EHF} = \frac{\bra \Phi_0 | \hat P^{\dag} \hat H \hat P | \Phi_0 \ket}{\bra \Phi_0 | \hat P^{\dag} \hat P | \Phi_0 \ket } =  \frac{\bra \Phi_0 | \hat H \hat P | \Phi_0 \ket}{\bra \Phi_0| \hat P | \Phi_0 \ket },\label{eq:EEHF}
\end{align}
is variationally minimized. This approach in particular is called variation-after-projection (VAP), not to be confused with  projection-after-variation (PAV), which has been widely used in quantum chemistry. $\hat P|\Phi_0\ket$ spans a large part of the Hilbert space, and thus is expected to  capture most of static correlation in a black-box manner, i.e., no active space is required. At the same time, for this reason, it has long been thought in quantum chemistry that the full spin-projection is computationally demanding and horribly complicated, even for PAV. Recently, Jim$\rm\acute{e}$nez-Hoyos {\it et al.},\cite{Jimenez12} however, have shown a feasible and clear way to accomplish VAP by using the spin-projection operator of the general integral form,
\begin{align}
{\hat P}^s_{mk} = |s;m\ket \bra s;k| = \frac{2s+1}{8\pi^2}\int D_{mk}^{s*}(\Omega) \hat R(\Omega) d\Omega, \label{eq:Ps}
\end{align}
instead of the famous L${\rm \ddot{o}}$wdin projector.\cite{Lowdin55} 
Here $s$ is the total spin, $m$ and $k$ are the spin angular momentum, $D_{mk}^s(\Omega) = \bra s;m | {\hat R}(\Omega)|s;k\ket$ is the Wigner matrix, and  $\hat R(\alpha,\beta,\gamma) = e^{i\alpha \hat S_z}e^{i\beta \hat S_y} e^{i\gamma \hat S_z}$ is a unitary rotation operator. With this formalism, the computational effort for EHF is known to be similar to that of mean-field methods. 

Although EHF efficiently describes static correlation, it neglects a vast amount of dynamical correlation, which is necessary for chemical accuracy. In order to remedy this, there have been recently extensive work attempting to incorporate the residual dynamical correlation into EHF, in the context of density functional correlation\cite{Garza13} as well as non-orthogonal CI,\cite{Rodriguez13} with promising results. In this work, we propose a perturbative approach based on MP2, which we shall hereby term extended MP2 (EMP2). Since MP2 correlation is almost exclusively of dynamical character, EMP2 should provide a seamless description of both static and dynamical correlation effects. A similar idea was pursued for PAV in the late 1980s,\cite{Schlegel86,Knowles88A} but was immediately abandoned due to its enormous computational cost even for approximate projection and many undesired features such as pronounced derivative discontinuities in potential energy surfaces. Below we show that, with the present scheme, all of these obstacles can be thoroughly resolved with full spin-projection.

{\it Theory.} 
Throughout this Communication, we restrict ourselves to the cases where $|\Phi_0\ket$ is an eigenstate of ${\hat S}_z$ but not of ${\hat S}^2$, i.e. an unrestricted HF type determinant, and thus $\hat P = \hat P_{mm}^s$. We will also adopt the conventional notations of orbital indices: $i,j,k,l$ for occupied, $a,b,c,d$ for virtual, and $p,q,r,s$ for all orbitals.  

Perturbation approaches for projected wave functions have been proposed by many others.\cite{Amos70,Peierls73,Knowles88A} Here we will derive our own scheme. We start by partitioning the Hamiltonian into $\hat H = \hat H_0 + \lambda \hat V$, such that 
\begin{align}
\hat H_0|\Phi_0\ket = E_0|\Phi_0\ket.\label{eq:E0}
\end {align}
We remind the reader that $|\Phi_0\ket$ is the broken symmetry deformed state. 
Given the Sch${\rm\ddot{o}}$dinger equation,
\begin{align}
\hat H|\Psi\ket ={\cal E} |\Psi\ket,\label{eq:SE}
\end{align}
$\cal E$ and  $|\Psi\ket$ are expanded around $E_0$ and $|\Phi_0\ket$ to find $n$-th order energies and wave functions, $E_n$ and $|\Phi_n\ket$.   The MP2 energy expression then becomes
\begin{align}
E^{(0)}_{\rm MP2} = \bra \Phi_0 | \hat H|\Phi_0 + \Phi_1 \ket  = E_{\rm HF}^{(0)} + E_{2}^{(0)}
\end{align}
where $E_{\rm HF}^{(0)}$ and $E_2^{(0)}$ are the HF energy and the second order perturbation correlation energy of the deformed state, and $|\Phi_1\ket$ is the first order wave function, which we will define later for our case. Note that we have not yet defined $\hat H_0$. Nevertheless, it is an independent particle symmetry broken Hamiltonian, and $|\Phi_1\ket$  consists of up to doubly excited determinants from $|\Phi_0\ket$.

Here our goal is to derive a perturbation theory that begins with $|\Psi_{\rm EHF}\ket \equiv \hat P|\Phi_0\ket$ and accomplishes the exact energy at the infinite order limit. Because $|\Psi_{\rm EHF}\ket$ has no well-defined independent particle Hamiltonian, however, one faces the difficulty of defining an appropriate zeroth order Hamiltonian. It should be clear that $\hat H_0$ defined in Eq.(\ref{eq:E0}) is not suitable, as $|\Psi_{\rm EHF}\ket$ is {\it not} its eigenstate. Hence,
we consider the expansion of $\cal E$ and $|\Psi\ket$ in the projected space around $|\Psi_{\rm EHF}\ket$. In the present scheme, our expansion for the wave function is based on the MP partitioning, given by
\begin{align}
|\Psi\ket &= \hat P |\Phi_0\ket + \lambda \hat P |\Phi_1\ket + \cdots. \label{eq:PPsi}
\end{align}
This is possible because the exact wave function can always be chosen as an eigenstate of $\hat P$, i.e., $\hat P|\Psi\ket = \Lambda|\Psi\ket$. Note that the spaces spanned by $\hat P|\Phi_n\ket$ are not orthogonal one another, and are necessarily overcomplete.\cite{Amos70,Peierls73} 
Eq. (\ref{eq:PPsi}) allows us to write the exact energy in the intermediate normalization,
\begin{align}
{\cal E}(\lambda) &= \frac{\bra \Psi_{\rm EHF} | \hat H \hat P | \Phi_0 + \lambda \Phi_1 + \cdots\ket}{\bra \Psi_{\rm EHF} |\hat P | \Phi_0 + \lambda \Phi_1 + \cdots\ket}\\
&= E_{\rm EHF} + \lambda {\cal E}_2 + \lambda^2 {\cal E}_3\cdots,
\end{align}
which achieves our goal, i.e., ${\cal E}(0)=E_{\rm EHF}$ and ${\cal E}(1) = {\cal E}$. All the perturbative information is then carried in Eq.(\ref{eq:PPsi}) and one is free from defining a zeroth order Hamiltonian. ${\cal E}_2$ parallels second order Rayleigh-Schr${\rm\ddot{o}}$dinger perturbation theory. Thus, we refer to it as the second order energy.
Expanding ${\cal E}(\lambda)$ around $\lambda_0 = 0$ and setting $\lambda = 1$, we find
\begin{align}
&E_{\rm EMP2} = E_{\rm EHF}+ {\cal E}_2,\label{eq:EMP2B}\\
&{\cal E}_2 = \frac{\bra \Phi_0 | (\hat H - E_{\rm EHF}) \hat P| \Phi_1\ket }{\bra \Phi_0 | \hat P | \Phi_0\ket}.\label{eq:E2B}
\end{align}
This formalism has various desired features. First, each term is rigorously defined by the magnitude of order parameter $\lambda$. 
 Second, there is no singles contribution from $|\Phi_1\ket$ due to the generalized Brillouin theorem,
 \begin{align}
\bra \Phi_0 |(\hat H - E_{\rm EHF}) \hat P a_a^\dag a_i | \Phi_0\ket = 0,\label{eq:GBT}
\end{align} 
when the EHF state is stationary, similar to the property in the conventional MP$n$ theory.\footnote{In fact, Eq.(\ref{eq:GBT}) is the EHF Fock matrix element defined as $\partial E_{\rm EHF}/\partial P^{(0)}_{ia}$ with normalization $\bra\Phi_0|\hat P|\Phi_0\ket$.} Last, and perhaps most importantly, the perturbation series are spin-projected at all orders, including $|\Phi_1\ket$. In fact, it can be shown that Eq.(\ref{eq:E2B}) may be seen as the fully spin-projected MP2 if $|\Phi_0\ket$ is the stationary unrestricted HF state. That is, by defining a projector onto the complementary space orthogonal to $|\Psi_{\rm EHF}\ket$,
\begin{align}
\hat O =1- \frac{\hat P| \Phi_0 \ket \bra \Phi_0 | \hat P}{\bra \Phi_0|\hat P | \Phi_0\ket} = 1- \frac{|\Psi_{\rm EHF}\ket \bra \Psi_{\rm EHF}|}{\bra \Psi_{\rm EHF}|\Psi_{\rm EHF}\ket},
\end{align}
 Eqs.(\ref{eq:EMP2B}-\ref{eq:E2B}) are elegantly rewritten as 
\begin{align} 
E_{\rm EMP2} 
&= \frac{\bra \Phi_0 | \hat H  \hat P | \Phi_0 + \hat O\Phi_1\ket}{\bra \Phi_0 | \hat P | \Phi_0\ket}. \label{eq:EEMP2B}
\end{align}
This clearly indicates that the last term in the numerator of Eq.(\ref{eq:EEMP2B}) lives in the space orthogonal to $|\Psi_{\rm EHF}\ket$ to eliminate double-counting of correlation effects, and is subject to spin-projection. We also note that the occurrence of double-counting is a natural consequence because, again, the basis Eq.(\ref{eq:PPsi}) is overcomplete.

Now we shall move our attention to the definition of ${\hat H}_0$ and thus $|\Phi_1\ket$ for EMP2. The performance of a perturbation theory critically depends on ${\hat H}_0$. 
Since we rely on the MP expansion of the deformed state, i.e. Eq.(\ref{eq:PPsi}), 
a physical choice for $\hat H_0$ is given by HF-like orbital energies evaluated from $|\Phi_0\ket$. They are indeed an appropriate candidate in view of Eq.(\ref{eq:EEMP2B}): EMP2 may be regarded as the full spin-projection of broken-symmetry MP2 for some special case. Also, this scheme is guaranteed to reduce to the regular MP2 when $|\Phi_0\ket$ is already a spin-eigenstate or $\hat P = 1$, providing a seamless connection. 

At the stationary state, $|\Phi_0\ket$ is not an eigenfunction of a sum of Fock operators, which is $\hat H_0$ of the conventional MP2. 
However, $|\Phi_0\ket$ is still a Slater determinant, and $|\Psi_{\rm EHF}\ket$ is invariant with respect to a unitary rotation among $|\Phi_0\ket$. Hence, we diagonalize the occupied-occupied ($oo$) block and virtual-virtual ($vv$) block of the deformed Fock matrix, defined as
\begin{align}
F^{(0)}_{pq} = h^{(0)}_{pq} + \sum_{rs} P^{(0)}_{rs} \bra pr||qs\ket,
\end{align}
with the deformed density matrix $P^{(0)}_{rs} = \bra \Phi_0 | a_s^\dag a_r|\Phi_0\ket$, and choose $\hat H_0 = \sum_p \varepsilon_p a_p^\dag a_p$ with $\varepsilon_p=F^{(0)}_{pp}$. The orbital basis of this particular choice has been referred to as semi-canonical orbitals in literature. 
Consequently, we will have not only doubles but also, potentially, singles contributions in $|\Phi_1\ket$,
\begin{align}
|\Phi_1\ket &= \sum_{ia} |\Phi_i^a\ket \frac{F^{(0)}_{ia}}{\varepsilon_i - \varepsilon_a} + \frac{1}{4}\sum_{ijab} |\Phi_{ij}^{ab}\ket \frac{\bra ij || ab\ket}{\varepsilon_i + \varepsilon_j - \varepsilon_a - \varepsilon_b}\nonumber\\
&= \sum_{ia} t_i^a |\Phi_i^a\ket + \frac{1}{4}\sum_{ijab} t_{ij}^{ab} |\Phi_{ij}^{ab}\ket,\label{eq:Phi1}
\end{align}
because $F_{ia}^{(0)}$ are nonzero in general. As mentioned above, however, all the singles contribution strictly vanish through $\hat P$ due to the generalized Brillouin theorem, Eq.(\ref{eq:GBT}). Thus we only require the second term. 

Finally, we discuss how one evaluates the projected coupling terms $\bra \Phi_0 | \hat H \hat P| \Phi_{ij}^{ab} \ket$ and $\bra \Phi_0 |\hat P| \Phi_{ij}^{ab} \ket$ that appear in Eq.(\ref{eq:E2B}). In practice, each term can be decomposed to a discretized grid integration as
\begin{align}
\bra \Phi_0 | \hat H \hat P| \Phi_{ij}^{ab} \ket &= \sum_g^{N_{\rm grid}} w_g \bra \Phi_0 |\hat H \hat R_g|\Phi_{ij}^{ab}\ket, \label{eq:Hijab}
\end{align}
where $w_g$ are the grid weights and $\hat R_g$ is the rotation operator defined earlier but for each grid point $g$.\cite{Jimenez12} The brute-force calculation of this term with the generalized Wick theorem\cite{RingSchuck} would require ${\cal O}(N^4)$ for each matrix element and therefore it gives rise to a total complexity of ${\cal O}(N^8N_{\rm grid})$ for all the double substitutions, which is intractable. To ameliorate the computational effort, we will take a couple of steps. 

Inserting the identity operator, $1=|\Phi_0\ket + \sum_{kc}|\Phi_k^c\ket  +\cdots$,  between ${\hat H}$ and $\hat R_g$ in Eq. (\ref{eq:Hijab}), we arrive at
\begin{align}
\bra \Phi_0 | \hat H \hat R_g| \Phi_{ij}^{ab} \ket 
=& E_{\rm HF}^{(0)}\bra \Phi_0 | \hat R_g | \Phi_{ij}^{ab}\ket  +\sum_{kc}F^{(0)}_{kc} \bra \Phi_k^c |\hat R_g|\Phi_{ij}^{ab}\ket\nonumber\\
&+\frac{1}{4}\sum_{klcd}\bra kl||cd\ket  \bra \Phi_{kl}^{cd} |\hat R_g|\Phi_{ij}^{ab}\ket\label{eq:Hgijab}.
\end{align}
In this way, only the rotation couplings $\bra \Phi_{kl}^{cd} |\hat R_g|\Phi_{ij}^{ab}\ket$, etc, are to be evaluated.

Here, the important realization is that $\hat R_g$ will {\it not} mix orbitals with one another, but instead independently rotate each spin orbital to give general spin orbitals (i.e., $\alpha$ and $\beta$ spins are mixed). Therefore, the ${\hat R}_g$ rotation on an excited determinant $|\Phi_{ij}^{ab}\ket = a_{a}^\dag a_{b}^\dag a_{j}a_{i} | \Phi_0\ket$ is {\it identical} to the corresponding excitation of the rotated determinant $|{^g\Phi}\ket \equiv \hat R_g|\Phi_0\ket$,
 \begin{align}
\hat R_g |\Phi_{ij}^{ab}\ket =c_a^\dag c_b^\dag  c_j c_{i} |{^g\Phi}\ket = |^g\Phi_{ij}^{ab}\ket,
 \end{align}
where $c_p^\dag$ and $c_p$ are the rotated creation and annihilation operators, $c_p^\dag = \hat R_g a^\dag_p \hat R_g^\dag$, etc, and $^g|\Phi\ket = \prod_{i} c_i^\dag |-\ket$ with $|-\ket$ being the bare vacuum. Then, the rotation couplings in Eq.(\ref{eq:Hgijab}) are realized as just the overlaps between excited non-orthogonal general HF (GHF) determinants. 

This fact is particularly useful for our purpose, because all the simplicities in HF determinants are still available for $|^g\Phi\ket$. Among the most important ones is the corresponding pair theorem.\cite{Amos61} One can biorthogonalize the orbitals of $|\Phi_0\ket$ and $|{^g\Phi}\ket$, $|p\ket$ and $|{^gq}\ket$, by performing a singular value decomposition of the $oo$ and $vv$ blocks of the overlap matrix ${^gS}_{pq} = \bra p | {^gq}\ket$. By the aforementioned theorem, which of course holds for GHF determinants, the resulting ${^g\bf S}$ matrix in the corresponding orbital basis is banded: not only the $oo$ and $vv$ blocks but also the $ov$ and $vo$ blocks can be chosen to be diagonal. This greatly simplifies the overlap evaluation\cite{Yost13} and makes it possible to retain only ${\cal O}(N^4)$ cost for the contraction of Eq.(\ref{eq:Hgijab}), using the significant sparsity of $\bra \Phi_{kl}^{cd}|{^g\Phi_{ij}^{ab}}\ket$ with a very simple algorithm. The limiting step of EMP2 is thus the computation and transformation of $t$ amplitudes as well as two electron integrals as in the regular MP2, which scales as ${\cal O}(N^5)$. Note that the final energy is invariant with respect to these orbital rotations.

\begin{figure}[t]
\caption{$Top.$ Potential energy curves of the H$_2$ molecule. $Bottom.$ Deviations from the FCI energy for FH.}\label{fig:FH}
\includegraphics[width=75mm]{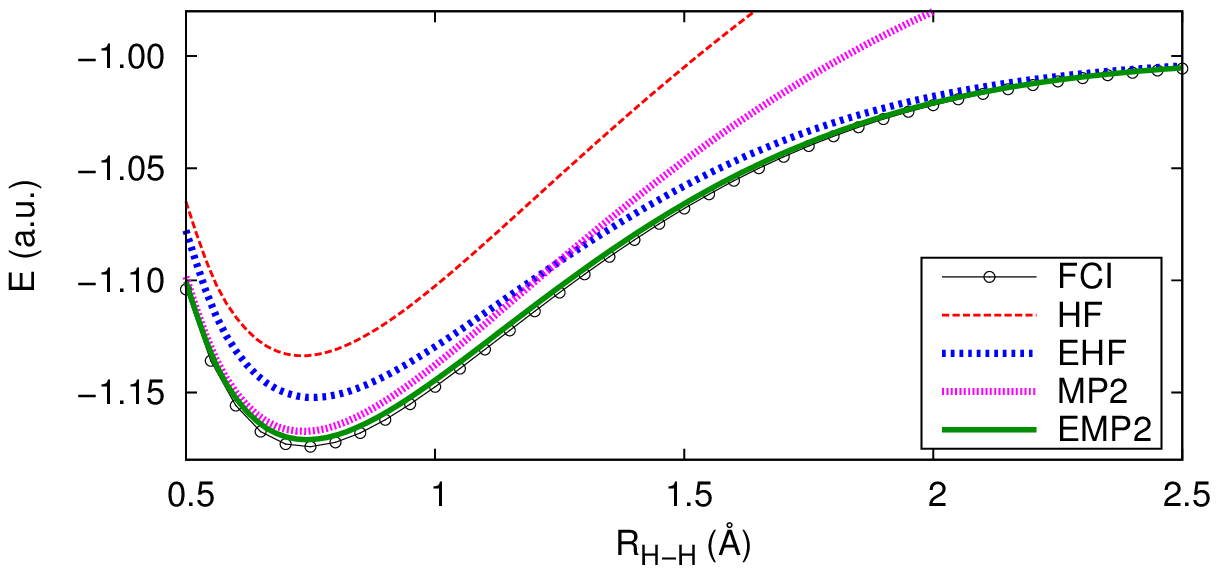} 
\includegraphics[width=75mm]{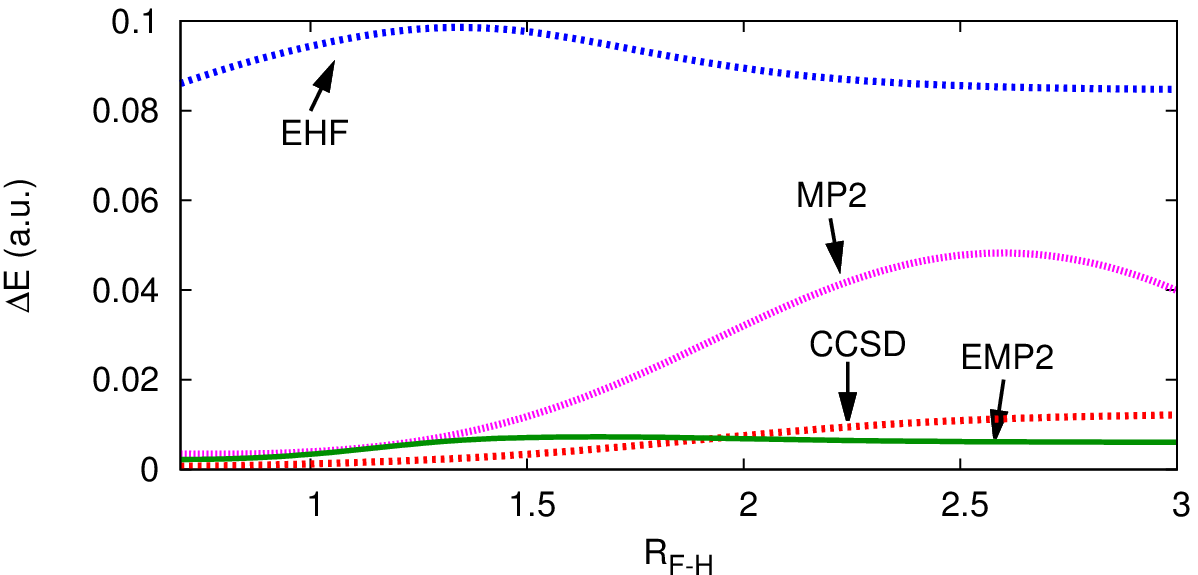} 
\end{figure}
\begin{table}[b!]
\caption{Non-parallelity error against FCI in kcal/mol.}\label{tb:NPE}
\begin{tabular}{lrrrrrrrr}
\hline\hline
	&	EHF	&		EMP2&	MP2	&CCSD\\
\hline
H$_2$\footnote{cc-pV5Z.} &4.5  & 0.7 & 13.3 & 0.0\\
FH&   2.8 & 0.8 & 10.2 & 2.4\\
H$_2$O  & 10.0&	2.2&	36.8&	7.3\\
N$_2$\footnote{1$s$ orbitals are frozen in FCI.} &	24.6		& 6.3	&439.7&	23.3\\
\hline\hline
\end{tabular}
\end{table}

\begin{table}[t!]
\caption{Singlet-triplet splitting energies for diatomic molecules in kcal/mol ($\Delta E_{\rm ST} =E(^1\Delta) - E(^3\Sigma)$).}\label{tb:EST}
\begin{tabular}{cccccccccc}
\hline\hline
&			EHF&		EMP2 &	MP2&	CCSD&	FCI\\\hline
NH&			49.60&		45.47&	58.06&	50.85&	45.51\\
OH${^-}$&	62.58&		58.06&	74.83&	64.46&	58.34\\
O$_2$\footnote[1]{1$s$ orbitals are frozen in FCI.}&		35.99&		28.80&	30.75&	32.71&	25.54\\
NF$^a$&			47.87&		40.21&	50.69&	48.48&	40.87\\
\vspace{-0.2cm}\\
MAE& 	6.44&		1.06&	11.02&	6.56\\
\hline\hline
\end{tabular}
\end{table}

{\it Results}.
We have implemented Eq.(\ref{eq:E2B}) in our in-house quantum chemistry program with the proposed contraction scheme. All the calculations were done with a 6-31G basis to enable the direct comparison with the exact full CI (FCI) results, except the hydrogen molecule. In the top panel of Figure \ref{fig:FH}, we depict the potential energy curve of H$_2$ with cc-$p$V5Z. As is well known, the MP2 energy is accurate in the short range where a tremendous amount of dynamical correlation is required, but it completely fails when a bond is stretched, due to its inability to describe static correlation. It is evident that the almost opposite event is observed in EHF. It dissociates H$_2$ exactly, being less accurate in the vicinity of the equilibrium bond length, R$_e$. As one would expect, EMP2 eliminates these disadvantages. It gives even slightly better energies than MP2 near R$_e$ where static correlation is considered negligible, while it starts to gain static correlation seamlessly toward the dissociation limit. Overall, the potential curve of EMP2 is in excellent agreement with FCI; the mean absolute error (MAE) is only 1.1 kcal/mol. These behaviors of correlation effects can be seen generically. The bottom panel of Figure \ref{fig:FH} shows the deviation of the total energy from FCI in the hydrogen fluoride molecule dissociation. Again, the conventional MP2  becomes notoriously worse after R$_{\rm F-H} = 1.5$ {\AA} due to the degeneracy appearing. While the EHF error is mostly flat and goes to the correct dissociation limit (although not size-consistent \cite{Jimenez12}), it vastly underestimates the dynamical correlation. CCSD, which almost superposes on FCI near R$_e$, loses its accuracy significantly and is usually difficult to converge as the bond is stretched. EMP2 yields the most accurate results over the entire region. The error observed throughout the dissociation coordinate is almost constant for this case. In Table \ref{tb:NPE}, we list the non-parallelity errors (NPE), defined as the error deviation from its MAE, i.e., NPE $={\rm avg} (|\Delta E - {\rm MAE}|)$, a measure of how parallel the potential energy curve is to FCI. We also performed the same analysis for the H$_2$O (symmetric dissociation) and N$_2$ molecules, all listed in Table \ref{tb:NPE}, showing the good performance of EMP2. 

\begin{figure}[t!]
\caption{Errors in potential energy curves of the $B{^1\Sigma}_u^+$  state in H$_2$.}\label{fig:H2_ex}
\includegraphics[width=80mm]{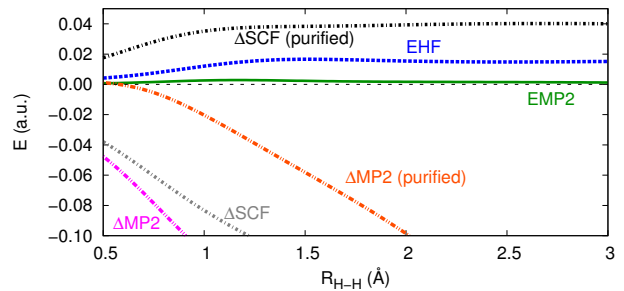} 
\end{figure}

We also report singlet-triplet splitting energies, $\Delta E_{\rm ST} $, of small diatomic molecules. The experimental geometries are used,\cite{CRC} and 1$s$ orbitals are frozen in the FCI calculations  for O$_2$ and NF. 
For triplet states, we have used unrestricted methods for MP2 and CCSD.
$\Delta E_{\rm ST}$ is only accurate if a method offers a balanced description of dynamical and static correlations. As tabulated in Table \ref{tb:EST}, we found EHF and CCSD share a similar quality with MAEs of 6.44 and 6.56 kcal/mol, respectively. Although CCSD includes the required (double) excitations, singlet states are not treated as accurately as are triplet states because the reference closed-shell HF orbitals are inadequate. This causes the consistent overestimation of $\Delta E_{\rm ST}$. On the other hand, EMP2 outperforms other methods, achieving an impressive improvement over EHF with a MAE of 1.06 kcal/mol.

Finally, we investigate the excited state of H$_2$ by $\Delta$SCF where an excited configuration is achieved by occupying electrons in virtual orbitals.\cite{Gilbert08} This state-specific non-aufbau approach is also applicable to EHF and allows us to compute low-lying excited states as a spin-pure state. Therefore, as opposed to the conventional $\Delta$SCF using HF, which suffers from significant spin-contamination, the spin-purification procedure is not needed in EHF. Furthermore, since such EHF state is stationary, one can directly perform EMP2. Figure \ref{fig:H2_ex} presents the error in potential energy curve of the first excited $B{^1\Sigma_u^+}$ state of H$_2$ against the FCI result, using 6-31G**. While $\Delta$SCF (i.e., HF) gives a qualitatively reasonable potential when purified, MP2 correction to $\Delta$SCF (denoted as $\Delta$MP2) miserably diverges. This is due to the degeneracy appearing in the dissociation limit with the second excited state. EMP2, however, has no such issue. It improves the EHF energy by adding dynamical correlation on top of it, and yields almost the exact potential curve. This encouraging result demonstrates the applicability of EMP2 to excited states.

We close our discussions by stressing once again that the method presented here achieves a black-box treatment of accurate dynamical and static correlation with a moderate computational effort similar to the conventional MP2.

The authors are grateful to Gustavo E. Scuseria for fruitful discussions. This work was supported by NSF (CHE-1058219).

\bibliographystyle{aip}

\end{document}